\documentclass[
journal,
twocolumn,
]{IEEEtran}

\usepackage{float}
\usepackage{caption}

\usepackage[utf8]{inputenc} %
\usepackage[T1]{fontenc}    %
\usepackage{hyperref}       %

\usepackage{url}            %
\usepackage{booktabs}       %
\usepackage{amsmath,amsfonts}       %
\usepackage{nicefrac}       %
\usepackage{microtype}      %
\usepackage{array}
\RequirePackage{etexcmds}
\usepackage{flushend}
\usepackage{color}
\usepackage{cite}
\usepackage{graphicx}
\usepackage{epstopdf}
\usepackage{subcaption}
\usepackage{multirow}
\usepackage[table,xcdraw]{xcolor}
\usepackage{amsmath,stackrel}
\usepackage{amsfonts}
\usepackage{mathalfa}
\usepackage{amssymb,mathtools}
\usepackage{amscd}
\usepackage{multirow}
\usepackage{makecell}
\usepackage{booktabs}
\usepackage[ruled,vlined,commentsnumbered,linesnumbered,lined,boxed]{algorithm2e}
\usepackage{algorithmicx}
\usepackage{algpseudocode}

\usepackage{ifthen} %
\usepackage{xifthen} %

\usepackage{xpatch}
\usepackage{blindtext}

\usepackage[colorinlistoftodos,prependcaption,textsize=tiny]{todonotes}

\usepackage{listings}
\usepackage{paralist}
\usepackage{enumerate}
\usepackage[shortlabels]{enumitem}
\newlist{abbrv}{itemize}{1}
\setlist[abbrv,1]{label=,labelwidth=1in,align=parleft,itemsep=0.1\baselineskip,leftmargin=!, font=\normalfont\large}

\usepackage{pgf}
\usepackage{tikz}
\usetikzlibrary{decorations.text}
\usetikzlibrary{backgrounds,shapes} %
\usetikzlibrary{arrows,automata}

\def\Rwnd{\textbf{\textit{Rwnd }}}

\SetKwProg{Fn}{Function}{}{}

\newcommand*{\affmark}[1][*]{\textsuperscript{#1}}

\newcommand\scheme{{\textit{{FairQ~}}}}

\newcounter{notebbcnt}
\title{FairQ: Fair and Fast Rate Allocation in Data Centers}

\author{Ahmed M. Abdelmoniem\affmark[1] \quad and \quad Brahim Bensaou\affmark[2]\\
\affmark[1]Queen Mary University of London, UK\\ \affmark[2]  Hong Kong University of Science and Technology, HK
}

\begin{document}

\maketitle

\begin{abstract}
The peculiar congestion patterns in data centers are caused by the bursty and composite nature of traffic, the small bandwidth-delay product, and the tiny switch buffers. It is not practical to modify TCP to adapt to data centers, especially in public clouds where multiple congestion control protocols coexist. In this work, we design a switch-based method to address such congestion issues; our approach does not require any modification to TCP, which enables easy and seamless deployment in public data centers via switch software update. We first present a simple analysis to demonstrate the stability and effectiveness of the scheme, and then we discuss a hardware NetFPGA switch-based prototype. The experimental results from real deployments in a small testbed cluster show the effectiveness of our approach.
\end{abstract}

\section{Introduction}

Cloud computing has dramatically risen in popularity in the past two decades. The idiosyncrasies of data center networks traffic brought back to the surface the old tug-of-war problem between mice flows and elephant flows \cite{GuoLiang2001,Wei2016}. In a nutshell, like many recent works, our goal in this article is to study, design and implement the means to reconcile the flow delivery timeliness requirements of the former with the high throughput requirements of the latter, under the stringent conditions imposed by data center network characteristics. Numerous congestion control mechanisms have been proposed for the Internet, high-speed WANs, lossy wireless networks and data centers, and the interested reader may refer to the following surveys for a comprehensive coverage~\cite{TCPsurvey, TCPfriendlysurvey, TCPwireless, TCPdatacenter}. Each algorithm aims to improve the way TCP reacts to congestion. 

The co-existence of various flows with different performance requirements ranging from synchronous mice flows to bulky elephant flows, poses another challenge to TCP in data center networks. This can be attributed to several issues, including the high bandwidth, low latency in data centers and the use of switches with shallow buffers. %
This mismatch between TCP's design philosophy and the characteristics of data centers leads to many complex non-conventional congestion events which are typically not observed in the Internet~\cite{Chen2009, Alizadeh2010, Wu2013}.%
~These events cannot simply be inferred from packet losses or duplicate ACKs, and hence they require special treatment. The typical TCP issues found in data centers are incast congestion and buffer-bloating which are defined as follows:
\begin{inparaenum}[1)]
\item \textbf{Incast congestion} occurs where many loosely time-correlated mice flows converge onto the same congested output port of a switch over a short period of time. With small buffers most of these flows would experience losses;
\item \textbf{Buffer-Bloating} occurs as a result of the normal stabilization process of TCP congestion control which requires the flows to {\it ``grab-all-bandwidth''} available, and as a result, ends up filling up the buffer (even though it is not necessary). Long-lived elephant flows occupy most of the buffer space persistently. In contrast, mice flows fail to grab their share of buffer and experience repeated losses, seeing at the end significantly long flow completion times.
\end{inparaenum}
These two problems may coincide due to the co-existence of flows of different natures and requirements competing for the same buffer of the bottleneck links.

Therefore, in this work, we take a flow-aware approach similar to traditional flow-based networks (e.g., the available bit rate service in ATM networks (ATM-ABR) \cite{ATM-ABR} or its Internet counterpart XCP \cite{Katabi2002}). The challenge that arises however is how to deploy such \textsl{flow-awareness in the flow-agnostic and flow-averse IP environment} without modifying the TCP sender and receiver algorithms or code. This requirement immediately disqualifies XCP, as it is a clean-slate redesign that requires not only changes to the routers but also to the sender and receiver. To achieve our goal, the switch must only be able to:
\begin{inparaenum}[\itshape i)\upshape]
\item track the number of active TCP flows per queue (instead of the full TCP state); 
\item calculate a fair share for each flow that traverses the output link; 
\item have the means to convey this fair share back to the traffic sources.
\end{inparaenum}
We accomplish this via a simple switch-based equal share allocation algorithm called \scheme\!\footnote{This work extends and builds on our prior work in~\cite{Ahmed-CLOUDNET-2015,Ahmed-IPCCC-2015}.} that:
\begin{inparaenum}[\itshape i)\upshape]
	\item uses simple counters to track the number of active flows for each output queue in the switch;
	\item modifies the switch to enable rewriting of the flow's fair share in the receiver window field of the TCP header as a means to enforce a maximum sending rate at the TCP source without changing the network stack of the end hosts.
\end{inparaenum}
This in the end amounts to deploying TCP flow control between the switch and the individual TCP sources that cross the switch with the objective of maintaining a target buffer occupancy. Since, TCP flow control is an integral part of any TCP incarnation, including XCP and DCTCP, the simple yet effective mechanism we propose would fit well without any change to TCP endpoints that may typically reside inside private virtual machines (VMs) or containers. Our contributions to this paper are: 
\begin{itemize}
\item We propose a novel TCP-invariant switch scheme, \scheme\!, to achieve fair-share allocations in data centers.
\item We mathematically model the \scheme\! dynamics and show the convergence of \scheme's flow control. 
\item We present software and hardware prototypes of \scheme in OpenVSwitch and NetFPGA platforms, respectively.\footnote{To help interested readers reproduce our results and for openness, we make the code and scripts of our implementations, simulations, and experiments available online at the following link: {\textbf{\url{http://github.com/ahmedcs/RWNDQ}}}.}
\item We evaluate \scheme\! via simulations and experiments in a testbed using software and hardware switches.  

\end{itemize}

The remainder of this paper is organized as follows, we first give a brief overview of the background and problems in Section~\ref{sec:problem}. Then, we discuss our proposed methodology and present the proposed switch queue management algorithm ``\scheme" in Section~\ref{sec:method}. We present the the hardware prototypes of \scheme\!  and its experiments in Sections~\ref{sec:hardware}.  Finally, we give the related work in Section~\ref{sec:related} and summarize our contributions in Section~\ref{sec:conclude}. We refer the reader to \cite{Ahmed-CLOUDNET-2015,Ahmed-IPCCC-2015,Ahmed-Arxiv-2021}, for details on the analytical model of \scheme\! showing its convergence and stability. And the evaluation of its performance via simulation and compare it to the alternative approaches as well as  the testbed experiments of the Linux kernel and Open vSwitch.

\section{Background}%
\label{sec:problem}

\subsection{The Intra-Protocol Unfairness in Data Centers}

The coexistence of TCP flows from the same variant should ideally result in a fair competition where each flow is able to grab an equal share of the bottleneck link capacity~\cite{Molnar2009}. This fairness is also known as the Round-Trip Time (RTT) fairness because it is conditional on the fairness of the RTT: that is, the TCP throughput is inversely proportional to the RTT, and two flows can nominally achieve the same throughput if they experience the same RTT~\cite{Gustavo2007}. In addition, since the Internet-centric design of TCP targets long-term fairness among competing flows, TCP's deployments in data center networks have inherited this goal. As a result, short-lived TCP flows that abound in data center networks do not last long enough to reach the steady state so they cannot obtain their nominal fair share, and often they experience losses leading to long Retransmission Timeouts (RTO). The short flows would benefit if short-term fairness is achieved but it is hard to achieve this for short-delay high-speed networks (e.g., data centers) with the current TCP design catered for long-delay low-speed networks (e.g., the Internet)~\cite{GuoLiang2001,Alizadeh2010}. %

\subsection{The TCP Flow Control mechanism}

To implement a flow-controlled byte stream reliable data transfer service on top of the Segmented TCP transmission, each TCP end-point reserves, during connection establishment, a buffer for receiving incoming segments from its peer. The main goal of this buffer is to simplify the implementation of a distributed flow control by simply ensuring that the sender never overflows the buffer space of the receiver. As such, the outgoing segments waiting for transmission or the arriving segments waiting to be consumed by the application are stored in the send buffer or receive buffer, respectively. To prevent receive buffers from overflowing, TCP provides the means for the receiver to pace the sender rate by controlling the extra amount of data that can be sent by the sender. This is achieved by returning a permissible ``window" of bytes with every ACK. %
TCP segment headers have a field named ``Receiver Window Size'' that serves the purpose of signalling the allowed number of bytes that the sender may transmit without overflowing the receive buffer~\cite{RFCTCP}. %
This field has a width of 16 bits thus allowing barely an increment of 64KB of data, which was sufficient in the early days of the Internet. Today, TCP includes an option called ``window scaling option'' 
{\cite{RFC7323}} that expands the field to 30 bits allowing windows of up to 1 GB.

\subsection{Relationship between Congestion Control and Flow Control}

In addition, to the flow control window, TCP congestion control also uses a window to enforce a calculated limit on the source sending rate based on the currently perceived congestion level. Even though the two mechanisms are considered to be different and target different purposes, functionally they are interrelated as they both are used to limit the TCP source sending rate. Generally speaking:
\begin{inparaenum}[1)]
\item \textbf{Flow Control:} adjusts the sending rate of the source to match the available buffer and processing speed of its peer;
\item \textbf{Congestion Control:} adapts the sending rate to the congestion state perceived from the network's implicit or explicit feedback.
\end{inparaenum}
At any instant in time, any TCP flow in the network is limited by either the remaining buffer space of its peer or the current value of its unused congestion window. TCP sets the relation between the congestion window $Cwnd$ and the receiver window $Rwnd$ as follows: $Swnd  = \min(Cwnd, Rwnd)$.

\subsection{The Role of Active Queue Management}

Active Queue Management (AQM) algorithms are deployed in switching devices to help control congestion by continuously monitoring the state of the output queues and taking an active role in relieving congestion if necessary. Typically the instantaneous (or average) queue size, arrival rate and/or departure rate are estimated and whenever their value exceeds a certain threshold the algorithm infers congestion on the link. Upon congestion, either, the algorithm proactively drops packets as a form of implicit congestion notification to the sources or it sends explicit congestion notification signals to the sources to adjust their sending rates. A typical example is the so-called RED AQM with Explicit Congestion Notification (ECN)~\cite{Floyd1993}.

\section{The Proposed Methodology}
\label{sec:method}

TCP is a full-duplex protocol where the two receiving end-points allocate a receiving buffer space to enable flow control operations. To this end, the receiver of one direction sends back to the sender ACK packets that are included in the header, in the 16-bit "Receive Window" field ($Rwnd$), the currently available buffer space. In high-speed networks, to prevent the protocol from becoming a bottleneck, the receiver window scaling option of TCP is invoked to semantically expand the 16-bit field up to 30 bits. The two end-points agree \textsl{a priori} on a number ($n$) between 0 and 14  that defines a multiplicative factor ($2^n$) applied to $Rwnd$ value to calculate the actual receiver window value. This allows for maximum receiver window values of up to 1 GByte when the scaling value is $n=14$. 
{RFC7323~\cite{RFC7323} states that the window scaling option can be set during the connection establishment phase in the SYN packet.}
In our approach, we propose to overwrite $Rwnd$ 
to indicate the bottleneck fair share of the bandwidth and buffer available for each flow sharing the same output port. As the ACKs traverse the switches in the reverse path towards the sender, each switch examines the ACK and modifies the $Rwnd$ value after adjusting it using the window scaling value if necessary. %

\subsection{System Design and Algorithm}

\small
\begin{table}[htbp]
	\caption{Variables and Parameters of \scheme\! Algorithm \ref{alg:rwndq}}
	\centering
		\begin{tabular}{|c|c|}
		\hline
		\textbf{Parameter name} 	& \textbf{Description} \\\hline
		$T$ 			& Timeout value for each window increment interval \\\hline
		$M$ 			& Number of increment intervals to wait for an update\\\hline
		$B$ 	         & Buffer size on the data path \\\hline
		$\alpha$ 		& Target level of queue occupancy	\\\hline\hline
		Variable name 	& Description \\\hline
		$Rwnd$ 		& Local receive window value for all flows\\\hline
		$\beta$     	& Number of current ongoing flows\\\hline
		$\gamma$ 	& Window increments of one update interval\\\hline
		$\Gamma$ 	& Counter of the number increments\\\hline
		$Q$ 		& Current output queue length in bytes  \\\hline
		$\kappa$    & The drift of $Q$ from the target $\alpha B$ \\\hline
		$P$		         & A packet\\\hline
		$Rwnd(P)$	& The value of receive window in TCP header\\\hline
		$Reserved(P)$	& The value of reserved bits in TCP header\\\hline
		$slow\_start$ & The current state of slow-start flag\\\hline
		\end{tabular}
	\label{tab:varAlgo1}
\end{table}
\normalsize

The main variables and parameters used in \scheme\! algorithm are described in Table~\ref{tab:varAlgo1}. Note that $T$, $M$ and $\alpha$ are parameters of the algorithm that can be chosen by the administrator. Algorithm \ref{alg:rwndq} is shown as a set of event handler functions in an event-driven environment. It runs on the switch and responds to two major events: packet arrivals, and timer-based local window update events.  

\textbf{Upon a packet arrival:} the algorithm updates the maximum packet size seen so far. If this is the first flow to arrive at this port, then the current window is initially set to the target queue worth of bytes then \scheme\! (optionally) enters the slow-start phase to start probing for the effective window size. This is because initially the end-to-end bandwidth-delay product is unknown to the switch and hence the available bandwidth has to be probed. Subsequently, for each new flow, the current window is divided equally among all flows. If the ACK bit is set, the receive window field $Rwnd(P)$ is rescaled by the scale factor. Then the rescaled receive window $Rwnd(P)<<scale$ is compared to the current local window value $Rwnd$ of the ingress queue. If $Rwnd$ is smaller than $Rwnd(P)$, then this packet is updated with the current local window $Rwnd$ after being scaled by the scale factor (i.e., $Rwnd>>scale$). Note that, in Appendix~\ref{sec:scale}, we show that the scale factor can be encoded into the reserved bits field (i.e., $Reserved(P)$) by a hypervisor-level shim-layer running on the physical machine.

\textbf{Upon window update timer expiry:} $\kappa$ is calculated to track the deviation of the current queue length from the target. The ratio controls the fraction of the Maximum Segment Size (MSS) added or subtracted from the current value of $\gamma$. After a number $\Gamma$ of such updates, the current value of the local per-queue window is updated. Then, if slow start is active, \scheme\! adds two MSS to the window, otherwise it adds the current value of $\gamma$ to the per-queue local window $Rwnd$. Notice that the value of the window increment is updated $M$ times before it is added to the actual value of TCP receive window $Rwnd(P)$ that is conveyed to the TCP sender. This enables a highly accurate estimate of the increment while keeping the number of receive window field $Rwnd(P)$ rewrites in the packet header reasonable.

The design of \scheme\! enables it to maintain a very low loss rate and to leave enough buffer space to absorb sudden traffic bursts while keeping links highly utilized. It embodies two principles, i) a congestion controller that adopts a proportional increase, proportional decrease approach where the window is expanded or shrunk in proportion to the severity of the congestion which is reflected by the backlog in excess of the target queue threshold; ii) a fairness controller that divides the amount of increase or decrease equally among all ongoing flows. This makes it appropriate to handle well the co-existence of mice and elephants. Each switch port is associated with a nominal window variable. Initially and whenever the number of ongoing flows drops to zero, the algorithm goes into the slow start mode, where the window is incremented by two MSS after the end of each update period. When the queue exceeds the target occupancy, the algorithm goes into congestion avoidance and the window is decremented in proportion to the backlog in excess of the target queue occupancy. 

\begin{algorithm}[h]
 \caption{Switch-based Equal Fair-Share AQM (\scheme\!) Algorithm}
\label{alg:rwndq}
\Fn{Packet Departure Event Handler ($P$)}
{	
	\lIf{$MSS \leq TCPSize(P)$}
	{
               $MSS \leftarrow TCP\_Size(P)$
     }
	\If{$SYNACK(P)$}
	{
	         \lIf{$\beta \leq 0$}
	            {
					$Rwnd \leftarrow \alpha \times B$
		     	}			
		     \Else
		     {
			  $Rwnd \leftarrow Rwnd \times \frac{\beta}{\beta+1}$\;
		     }
		    $\beta \leftarrow \beta + 1$\;
     }
	\If{$FIN(P)$}
	{	
		 $\beta \leftarrow \beta - 1$\;
		\lIf{$\beta \geq 0$}
		{
			$Rwnd \leftarrow Rwnd \times \frac{\beta+1}{\beta}$
		}				   
		\Else
		{
			  $Rwnd \leftarrow \alpha \times B$\; 
			  $slow\_start \leftarrow True$\;			  
		}
	}
	\If{$ACK(P)$ \&\& ($Rwnd \leq Rwnd(P)<<Reserved(P)$)}
	{
		$Rwnd(P) \leftarrow Rwnd>>Reserved(P)$\;
	}
}
\Fn{Window Update Timer}
{
          $\kappa \leftarrow 1 - \frac{Q}{ B \times \alpha}$\;
          $\gamma\ \leftarrow \gamma  + \frac{\kappa \times MSS}{M}$\; 
          $\Gamma \leftarrow \Gamma + 1$\;
	      \If{$\Gamma == M$}
	      {
			 \lIf{$slow\_start == True$}
			 {
				$Rwnd \leftarrow Rwnd  + 2 \times MSS$
			}				
	        \lElse
	        {
				$Rwnd \leftarrow Rwnd  + \frac{\gamma}{\beta}$
	        }
	        \lIf{$Q \geq \alpha * B$}
			{
	                    $slowstart \leftarrow False$
			}
		    $\gamma \leftarrow 0$; $\Gamma \leftarrow 0$\;
	    }
}
\end{algorithm}
\normalsize

\subsection{Practical Aspects of The System}
\label{subsec:pracrwndq}

\textbf{Flow Tracking: } In principle \scheme\! is very effective in solving the problem of congestion, and actually avoiding it outright.
However, to enable its successful practical deployment, the following requirements need to be met:
\begin{inparaenum}[\itshape i)\upshape]
\item the ACKs must travel back along the reverse path taken by the corresponding data packets, as they are used for switch signalling,
\item the switch must be able to track the number of ongoing flows to enable fair sharing; and,
\item as the Rwnd field of the TCP header is used for signalling, the algorithm must take into the possible use of the window scaling option for each ongoing flow to avoid semantic mismatches between the receiver and the switch in terms Rwnd values.
\end{inparaenum}
To achieve the first requirement (i), two approaches are possible: either implement flow-aware routing in the open source network OS of the bare-metal switch, or, more likely, since SDN-based switches are much more common nowadays in data center, one can rely on the functions already provided by SDN to setup flow-based routing. To fulfill the second requirement (ii), one can implement a SYN/FIN based counting using hardware registers in the switch can be invoked, or again relying on an SDN controller to track the number of active flows via a special OpenFlow rule SYN/FIN packets; we implemented and tested the first approach in a NetFPGA platform and deployed the second in an SDN-enabled testbed. To meet the third requirement (iii), we can rely on assistance from the end hosts as described in greater detail hereafter.

\textbf{Role of SDN: } In the SDN approach, SDN capability to track flows, flow statistics and the scaling value sent in the SYN segments can be easily invoked to address the three requirements above. In contrast, if SDN is not available and the network elements are not prone to upgrade, additional knowledge of the DCN architecture and routing can enable the DCN operator to easily deploy \scheme\!. For example, if single-path routing is used, the learning ability of the switches can be invoked to implicitly assume that forward and reverse paths are already the same. If Equal-Cost Multi-Path (ECMP) routing is used a simple modification to the \scheme\! algorithm to equally divide the flow fair share among the multiple routing paths is easily applied. In addition, to track the number of active TCP flows, we can simply implement efficient header filters to track SYN/FIN flags for connection establishment or tear-down without per-flow state by using per-port hardware registers.  

\textbf{TCP Window Scaling: } The TCP window scaling option remains an important issue. In practice, this option is supposed to be activated to deal with long-fat pipes by increasing the receiver window from 64KB per flow to up to 1GB per flow. However, even in current low-latency DCNs with 10-100 Gbps interfaces, the scaling remains a necessary element and the receive window still needs to be scaled to maintain full link utilization. Even though, one could argue that the chances of having a single flow active on a given port is close to nil (considering the average number of flows per server in a private DCN measurement is about 36~\cite{Alizadeh2010}), a robust technique should provide the ability to rescale receive window values. According to the RFC~\cite{RFC7323}, the window scaling option is supposed to be negotiated between the sender and the receiver, and to enable it, both sender and receiver must send the window scaling option in the SYN and its corresponding SYN-ACK. However, in practice, the scaling value is not negotiated as different TCP implementations adopt different default values for scaling factor. For example, by default in MacOS the scaling exponent is set to three while Linux calculates it according to the allocated receiver buffer size. Furthermore, these values can be reconfigured by the application to be from 0 (i.e., up to 64KBytes for no-scaling) to 14 (i.e., up to 1 GBytes with scaling). To avoid any cognitive mismatch between the values set by \scheme\! in the receive window field and those interpreted at the receiver and to operate regardless of the link speed used in modern data centers, the following are the possible ways to solve this issue without modifying TCP in the VM:
\begin{itemize}
	\item  If the scaling option is negotiated then we propose to simply unify the value supported in the DCN by rewriting it in the SYN and its corresponding SYN-ACK during the phase of TCP connection establishment via an SDN rule in the switches or directly by \scheme\!; 
	\item However, if the TCP implementation informs the peers of the scaling value during connection setup, then we propose to have a lightweight shim layer at the end-hosts. This shim-layer tracks the per-flow scaling factor, recomputes the receive window of outgoing ACKs and resets the value to a pre-set network-wide scaling factor which is already configured on all the switches. 
	\item The last possibility, which we have adopted in our prototype (see Section~\ref{sec:hardware}), is to deploy a module or shim-layer in the hypervisor or server. The module collects the scaling factor used by each side which is typically propagated to the other side with the SYN. Then, the module uses four of the available reserved bits in the TCP header to encode the scaling factor and the switch uses this value to adjust the new receive window whenever it is updated. Hence, the resulting system does not require any changes to TCP and is transparent to VM's guest OS.
\end{itemize}

\textbf{Processing Complexity: } In terms of processing complexity, \scheme\! is a very simple algorithm with very low complexity and can be integrated easily into switches or routers. For example, it can be implemented in Linux-based routers as a module using the NetFilter framework which allows for modifications to the packet headers prior to their forwarding by the IP layer. This requires $O(1)$ computation per packet. \scheme\! can also cope with Internet checksum recalculation easily and efficiently after header modification, by applying a straightforward one's-complement add and subtract operations on the following three 16-bit words \cite{RFCTCP}: $CSum_{new} = CSum_{old} + Rwnd_{new} - Rwnd_{old}$. In addition, since \scheme\! is designed to deal with TCP traffic only, tracking the number of flows can be achieved in a scalable manner by monitoring SYN/SYN-ACK and FIN/FIN-ACK bits. This also requires $O(1)$. If necessary, disabling or unifying the window scaling factor in the switch also requires $O(1)$. All in all, all the operations required by the algorithm are $O(1)$ and most importantly involve only the switches/routers under the control of the DC operator. In particular, no modification to the TCP sender or receiver algorithms is needed. 

\textbf{Overhead of \scheme\! Deployment:} To reduce the overhead which \scheme\! imposes on core switches, its algorithm can be slightly modified by making use of one of the currently unused three flag bits in the TCP header or the high-order bit of the currently unused fragment offset in the IP header. This bit can be set on a predetermined interval or when the congestion level approaches a certain threshold as a flag on the ACKs that were modified by the intermediate routers. Then, when such scheme is applied to \scheme\!, only the hypervisor, ToR switches and/or lightly-loaded gateway/ingress routers are modified in such a way to store the current up-to-date per-flow receive window value and they become responsible for updating every ACK heading back to the sender. Consequently, by adopting this approach, there is no more extra burden on the intermediate routers or core switches to be involved in updating every ACK passing through them.

\textbf{Effect on Internet-facing TCP connections:} It is worth mentioning that, the WAN connections of data centers to the Internet in most cases are facing intra-Datacenter load balancers and proxies that split the TCP connection. Hence, TCP connections inside the data centers are effectively separated from the TCP traffic from outside the Internet. This avoids the possible issues that \scheme\! updating the receive window of the Internet-facing TCP connections which runs in a high Bandwidth-Delay Product (BDP) environment like the Internet.

\section{Deployment of The Hardware Prototype}
\label{sec:hardware}

DCN operators use commodity Ethernet switches with small buffers for interconnecting the servers mainly because of their low cost and ease of deployment. Typically, the in-network switches are under the control of the cloud (or data center) operator. These switches have a certain lifetime (and even worse they frequently fail long before that) after which they need to be replaced just like any other hardware equipment. Cloud operators are then forced to buy new (possibly upgraded models) switching devices to replace the old or failed ones. Hence, we believe that solutions involving very simple modifications of the switching chips are quite viable and appealing to cloud operators. Switching chip manufacturers would be motivated by marketing efficient data center-tailored switching chips that involve minimal hardware/software updates yet they provide significant performance gains. 

Our motivation towards building a proof-of-concept prototype is to provide a show-case for our proposed AQMs to the industry and potential switch fabric manufacturers. The prototype would also allow us to experiment with our AQM schemes in a realistic data center-like environment~\cite{Ahmed-ITCE-2019,Ahmed-INFOCOM-2019}. It would also allow for macro and micro benchmarking our schemes to establish its complexity, performance and overhead. This task was eased by the introduction of the 4-ports NetFPGA 1Gbps and 10Gbps cards. This project was initiated by Stanford University and the initial design and model of NetFPGA board was introduced in 2007. The aim of this project is to facilitate research and prototyping of networking protocols for routing and switching devices~\cite{Lockwood2007}. Hence, in this appendix, we introduced our prototyping procedure, discussion of practical findings and a summary of possible development methods in NetFPGA platform. Our sole aim of this appendix as a computer scientist is to ease the adoption of hardware prototyping for fellow computer scientists. Typically, computer science researchers lean away from such tedious tasks involving implementations using a non-familiar hardware language.  

Since switches are networking elements that are under-control of DC operators, they are typically a viable update option for DC operators. Hence, we present the prototyping of the hardware switch AQM designs of \scheme\!. The prototyped switch should possess the following features:
\begin{inparaenum}[(F1)] 
\item Improve latency-sensitive applications Flow Completion Time (FCT) and mitigate incast; 
\item Elephant flows can enjoy high sustained throughput in a work-conservative manner;
\item No modifications are imposed on the existing TCP protocol stack or anything that is controlled by the tenant\footnote{If changes to existing systems are needed (i.e., inevitable) then they must be in network devices and/or hypervisors that are fully under the control of the DCN operator};
\end{inparaenum}

To this end, we present ``FairSwitch'' resembling \scheme\!, the equal-share allocation scheme. In FairSwitch, the switch actively enforces rate adjustments in fair manner to regulate the queue at a certain target occupancy. The design leverages TCP flow control mechanism to control TCP flow rates by adjusting the ``advertised" window field without intervening with their current ``congestion" window update function.

\begin{figure}[h]
	\centering
		\includegraphics[width=1\columnwidth]{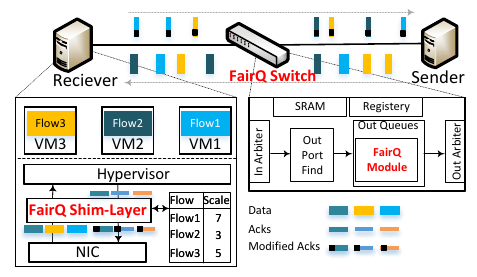}
	\caption{FairSwitch System: it consists of an end-host module that attaches scaling value to ACKs and FairSwitch which performs equal fair share allocation}
	\label{fig:system}
\end{figure}

We inspect the performance of the hardware prototype of \scheme\! called ``FairSwitch'' as well as its end-host helper module. \figurename~\ref{fig:system} shows the deployment of FairSwitch in a data center or cluster setup and how would its components interact with each other. The FairSwitch updates receive window of all incoming ACKs to assign a fair share of output port buffer. The scale factor written in the reserved bits of the packet header is used to rescale the new window so that it can be interpreted correctly by the ACK receiving end-point. \figurename~\ref{fig:system} shows that the end-host module tracks the scaling factor used by the local communicating end-points and explicitly appends this information only to the outgoing ACKs of the corresponding flow. The switch monitors the queue occupancy of the per-port output queue and the arrival of special packets such as SYN-ACK/FIN-ACK. The switch actively calculates a local per-port window value based on the amount of deviation from the target queue occupancy. The local window value is re-calculated each time a new flow is observed or an existing flow leaves the queue. The switch is also continuously rewriting the receive window field of all incoming ACKs if the local value is less than the value in the header.

\subsection{The NetFPGA Hardware Platform}
The NetFPGA is an open platform for gigabit Ethernet switching and routing that has been developed at Stanford University. Stanford University started the design and development of NetFPGA board in 2007 to facilitate research and prototyping of networking protocols for routers and switches \cite{Lockwood2007}. NetFPGA is a complete network hardware implemented on a FPGA along with 4 network ports. It is mostly used to build and test new protocols and network functions. The NetFPGA-1G board is equipped with Xilinx Virtex-II Pro 50 FPGA chip, 4 Gigabit Ethernet networking ports, Cypress 4.5 MB SRAM, Micron 64 MB DDR2 DRAM, and standard PCI Card \cite{netfpgaspec}. The board comes with multiple online open-source reference projects all written in Verilog and many other contributed open-source projects published by researchers. The NetFPGA platform has gained a lot of attention and was improved over the years to keep up with the new hardware configurations. At the time of this writing, the card we used is deprecated and replaced with 3 other flavors (i.e., SUME, CML and 10Gbps) for 1Gbps and 10Gbps networks with various pricing and educational discounts\footnote{NetFPGA website is available at http://www.netfpga.org}. All of our switch designs relied on the reference switch model whose image bitfile and source code are provided with the board and/or available online.

\subsection{The Reference Router and Switch Design}

\begin{figure}[ht]
	\centering
		\includegraphics[width=1\columnwidth]{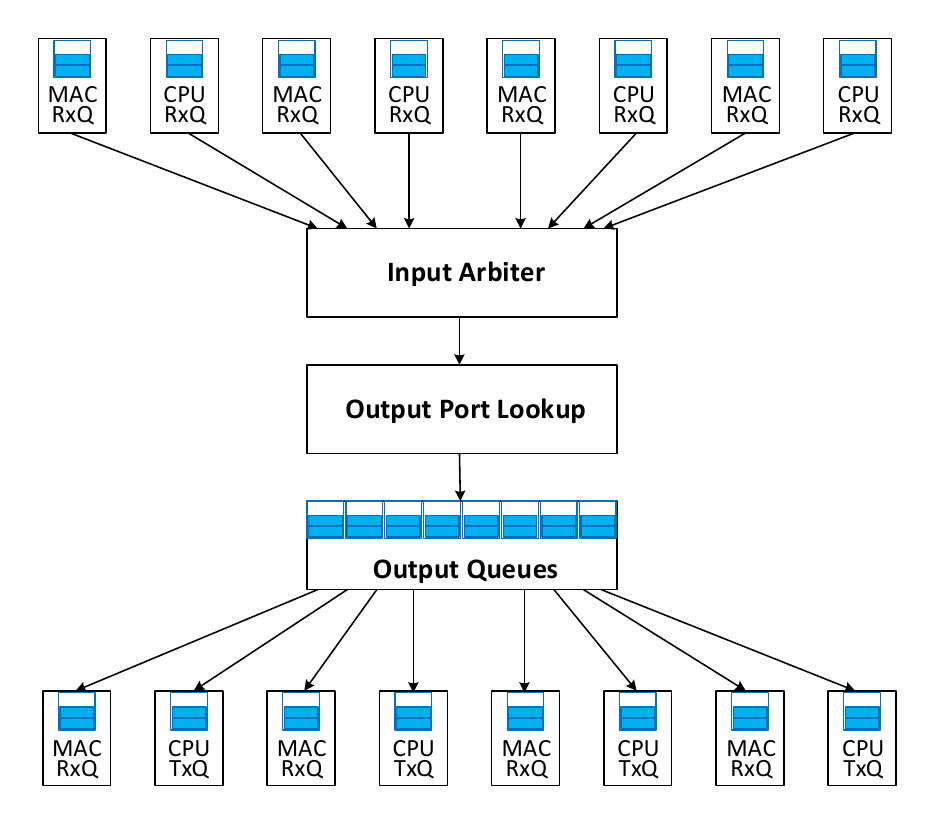}
	\caption{NetFPGA-based Modular Pipelined Design}
	\label{fig:netfpgapipeline}
\end{figure}

The reference IPv4 router and the switch designs are part of the NetFPGA base package. They consist of a number of Verilog modules that are all connected together in a pipelined manner Typically, they are organized in a way that allows the board to operate at the speed of 125 MHz (i.e., 8 nano-seconds per clock cycle)~\cite{Lockwood2007}. For the purpose of this appendix, first, we need to introduce the modular design of the NetFPGA to fully grasp its operational internals. Afterwards, this design is modified to implement our proposed switch AQM designs. \figurename~\ref{fig:netfpgapipeline} shows the pipelined structure of on which all the NetFPGA reference designs are built. The design features 4 physical (1 Gbps or 10 Gbps) Ethernet ports controlled by the FPGA board. Each port has an ingress or egress direction which is represented as a TX port (egress) and a RX port (ingress) for the switch. The FPGA exposes another 4 ``logical'' DMA ports (i.e., they are connected to the CPU of the hosting PC). Each logical port relates to each physical port to act as the channel between the physical port and the host on which the PCI card resides. In cases when the reference NIC is not in use,  the reference router/switch transfers the incoming packet to the respective DMA port and then to the CPU for processing whenever it unable to process or handle it.

Most of the major processing in the NetFPGA is carried out by 3 other modules that constitute the reference design. The first module is the input arbiter which uses a round-robin arbiter to pick up packets from the input queues and forward them to the corresponding FIFO of the output port Lookup module. The output lookup module does the packet processing, namely, parsing the Ethernet header, the IP/LPM lookup, and destination filtering and finally, modifying the Ethernet headers. It then hands the packet to the ``output queues'' module that initially stores the packet in the SRAM and then hands it to the corresponding physical port when it is available for transmission. In addition, between all the modules, the design employs an input FIFO in the pipeline between two successive modules (i.e., the output lines of the previous to the input lines of the next). This input FIFO is used for pipelining the design. The current module reads the data from the input FIFO as and when required and processes it. 

\subsection{An End-host Helper Module for Window Scaling}
\label{sec:scale}

All switch designs discussed in the following sections rely on a scale factor to rescale the used window for updating the ACKs when the receive window \Rwnd is updated. TCP specification \cite{RFC7323} states that the three-byte Scale option may be sent in all packets or only in a SYN segment by each TCP end-point to let its peer know what factor it uses for its own window value scaling. TCP implementations in most popular OSes including Linux adopt the latter approach to save overhead and wasted bandwidth of the former approach. The scaling may be unnecessary for networks with Bandwidth-Delay Product (BDP) of 31.25 Kbyte (i.e., C=1 Gbps and RTT=250$\mu$s). However, with the introduction of high speed links of 40 Gbps (i.e., BDP=1.25 Mbyte) and 100 Gbps (i.e., BDP=3.125 Mbyte), the scaling factor becomes necessary to utilize the bandwidth effectively. To this end and to avoid any expensive flow-level state tracking at the switch, we propose the adoption of a light-weight end-host (hypervisor) shim-layer to explicitly send scaling factor with outgoing ACKs. It extracts and stores from outgoing SYN and SYN-ACK packets the advertised scaling factor (i.e., within the window scaling option) for each established TCP flow. The system, at connection-setup, hashes the flows into a hash-table. It identifies the flows through the 4-tuple (i.e., source IP, dest. IP,  source port and dest. port) as the key and stores the window scale factor as the entry value. Flow entries are cleared from the table when a connection is closed (i.e., FIN is sent out). Then, %
the module writes the scale factor for all outgoing ACK packets in the 4-bit reserved field of TCP headers (alternatively, by using 4-bits of the receive window field and using the remaining 12 bits for window values)~\cite{RFCTCP}. The used reserved bits is cleared after their usage by the FairSwitch switch to avoid the packet being dropped by the destination due to an invalid TCP check-sum value which avoids the need for recalculating TCP checksum at the end-host and the switch. Typically, the module resides right above the NIC driver for non-virtualized setups, right below the hypervisor to support VMs in cloud data centers or implemented in the data-path of the NIC (e.g., FPGA-based NICs) if applicable. Hence, this placement does not touch any network stack implementation of the guest OS, making it readily deployable in production data centers.

\subsection{FairSwitch for Switch-based Fair Share Allocation}
\label{sec:fairswitch}

In this section, we discuss the hardware design aspects of FairSwitch in NetFPGA platform. FairSwitch is a hardware realization of a simple switch AQM mechanism that is based on the following ideas:
\begin{inparaenum}[\itshape 1) \upshape] 
\item we conclude that to govern the switch buffer usage among competing (distributed) TCP flows, a mechanism has to intervene to equally share the limited buffer space among these flows;
\item To be able to do so, the switches are the ones that have the most accurate information on the nature of congestion, therefore FairSwitch adopts an AQM namely ``\scheme\!'' to control buffer occupancy at a certain level;
\item To avoid modifying existing protocols at the sender and receiver, FairSwitch must rely on universally accepted mechanisms for congestion notification. The switch can be involved in  actively rewriting the \Rwnd field as a means to throttling the sender's rate of each connection.
\item How frequently should such rewriting take place? in this case, the switch rewrites \Rwnd for every ACK on the backward path to the senders. 
\end{inparaenum} 
Based on these ideas, FairSwitch is a hardware realization of the \scheme\! algorithm presented in Algorithm \ref{alg:rwndq}. \scheme\! is meant to equalize the \Rwnd field among TCP flows to achieve a target $Q\_Target$ occupancy that can reduce the latency for mice TCP flows and overcome the burstiness of elephant flows.

\subsection{NetFPGA Design and Implementation}
\begin{figure}[ht]
	\centering
		\includegraphics[width=1\columnwidth]{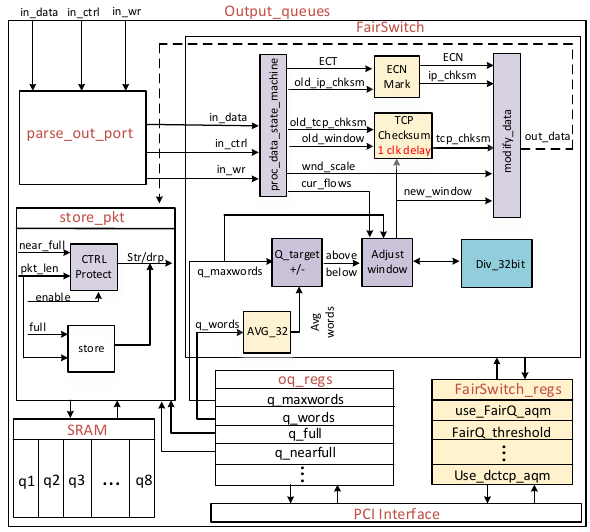}
	\caption{FairSwitch NetFPGA-based system design}
	\label{fig:fsnetfpga}
\end{figure}

Hardware implementation of FairSwitch was developed on the NetFPGA-1G card using ``Verliog" hardware description language.  For FairSwitch, we have chosen to build it on top of the reference switch project by integrating our module into the ``output queues" module. %

\subsubsection{NetFPGA FairSwitch module}
 
The main design is implemented in ``Fair\_Switch" module which is instantiated within the top ``output\_queues" module as shown in Fig~\ref{fig:fsnetfpga}. The Fair\_Switch module instantiates a number of helper modules namely ``avg\_32", ``div\_32", ``TCP\_checksum" and ``ECN\_mark" for different data processing. The main inputs to the module are the incoming packet (i.e., in\_data), control information added by the NetFPGA module (i.e., in\_ctrl) and data ready signal (i.e., in\_wr) all coming from the output\_queues module and the main outputs are the (un)modified packet information (i.e., out\_data, out\_ctrl and out\_wr) to be written into a FIFO for later storage in the destination queue's SRAM space. Other worth-mentioning inputs are q\_maxwords and q\_words coming from oq\_regs module which is the allocated SRAM space for the destination queue and the number of words currently used by stored packets. 

The module keeps track of separate per-output-queue counters for SYN/SYN-ACK, FIN/FIN-ACK and RST packets seen going through the queue. This information can be extracted by applying the state machine on in\_data to process the headers. Typically, when line in\_wr is active and line in\_ctrl is disabled then the word under processing is part of either the header of the packet or the meta-data added by NetFPGA. The current number of active flows is increased by one for the incoming and outgoing queue, each time a SYN-ACK is received. Whenever a FIN-ACK/RST is received, this number is decreased by one for the outgoing direction of the FIN-ACK.  At any time, the absolute difference between SYNs and FINs represents the number of currently active flows. This value is reset whenever no data packets is received from any of the active flows for enough period of time (i.e., 1 sec). The decision that the safe threshold (i.e., target queue occupancy) has been exceeded is evaluated for every output-queue. This is achieved by the current average number of occupied words, and the maximum number of words in the queue right-shifted by the target threshold\footnote{Right shifting by x is equivalent to multiplying by $2^{-x}$}. If the queue occupancy is above or below the threshold, the current queue local window is readjusted following the \scheme\! Algorithm~\ref{alg:rwndq}. Typically, the new window value is re-calculated, whenever the variable cur\_flows is updated due to a change in the number of currently ongoing connections. Then, in this case, the target occupancy (i.e., q\_maxwords $\geq$ rwndq\_threhshold) is divided by cur\_flows to calculate the new window value. And then, the window is adjusted up and down to maintain the target threshold.

On the arrival of each ACK, before it is forwarded to a certain output queue (port), the receive window field of TCP header is updated (only if it is less than the current value in TCP header) with the newly calculated window. The new window is right shifted by the ``window scale'' value which can be extracted out from the reserved-bits field~\cite{RFCTCP} %
Note that, the rwndq\_threshold should be set in a way to preserve an almost empty queue for absorbing elephants and large enough to keep the link busy by elephants.

\subsubsection{FairSwitch helper modules}

FairSwitch module relies on the following helper modules instantiated within to perform various functions: 
\begin{enumerate}
\item \textbf{AVG\_32}: This module is used to calculate a running average of the number of words used per-output-queue. This is accomplished by averaging q\_words over the last 32-samples taken every avg\_queue\_time. avg\_queue\_time is a customizable timer (default set to 48$\mu$s which equals the transmission time of four 1500 Bytes sized packets over 1 Gbps link). 
\item \textbf{TCP\_checksum}: This module calculates the new TCP checksum using the incremental update~\cite{RFCchecksum} method shown in line 13 of Algorithm~\ref{alg:rwndq}. It takes the window, new window and current checksum as input and outputs the new checksum value in the next clock cycle. For this reason, the in\_data needs to be delayed by one clock cycle to be able to write it into the correct word of the out\_data.
\item \textbf{ECN\_mark}: This module implements the DCTCP's AQM marking mechanism, in which packets are marked whenever a customizable fraction (default is 25\% which is close to the recommended 20\% value) of the maximum queue occupancy has been exceeded. It also recalculates a new IP checksum using ECT (01 or 10), ECN (11) and the old checksum value using the incremental update~\cite{RFCchecksum} method. In this case, however, no extra delay is needed as IP checksum is one word past the TOS field\footnote{NetFPGA processes data in words where each word consists of 64-bits (i.e., 8 bytes) and takes 1 clock cycle.}.
\item \textbf{DIV\_32bit}: This module is used for the window calculation by dividing the left-shifted (i.e., $gg$) value q\_maxwords by rwndq\_threshold over cur\_flows (i.e., setting a window of equal buffer share among flows). This divider is generated from Xilinx IP cores library and an instance of it is invoked. The division process takes 18 clock-cycles to complete but there is no need to add an extra 18 cycle delay to in\_data. This is because if this feature is used, the newly calculated window is updated with SYN/FIN arrivals and ACKs arrive with $\approx$ half RTT from SYN's arrival in the backward path.
\item \textbf{FairSwitch\_regs}: This module is instantiated in the output\_queues module to store all FairSwitch-related parameters (stored in registers) for usage by various module functions. It also communicates with the Peripheral Component Interconnect (PCI) interface for reading and/or setting the FairSwitch-related parameter values.
\end{enumerate}

\figurename~\ref{fig:fsnetfpga} also shows how FairSwitch module interacts with the other modules within output\_queues main module. Specifically, after the packet is processed by output\_port\_lookup module to determine the outgoing port, it is processed and stored by output\_queues for later transmission by the input\_arbiter (or output\_arbiter in this case) as illustrated in \figurename~\ref{fig:netfpgapipeline}. The destination port information is first parsed and extracted by parse\_out\_port module from the control headers which is added by output\_port\_lookup. Then the packet is processed by Fair\_Switch module and its output (i.e., out\_data) is sent to store\_pkt module for storing it in the SRAM or dropping it otherwise.

\begin{table}[ht]
\caption{Comparison of resource usage between the reference switch and FairSwitch}
\centering
 \resizebox{\columnwidth}{!}{
		\begin{tabular}{|c|c|c|c|c|c|}
		\hline
		 {}	                & \multicolumn{2}{c|}{Reference Switch} &   \multicolumn{2}{c|}{FairSwitch} & NetFPGA Total \\\cline{2-5}
		 {}		              & Used & Percentage & Used & Percentage & {} \\\hline
		Slices   & 9605 & 40\% & 10851 & 45\%	& 23616\\\hline
		Slice Flip Flops    & 8364 & 17\% & 9419 & 19\% &  47323	\\\hline
		4 input LUTs  & 14067 & 29\% & 15985 & 33\% & 47232 \\\hline
		BRAMs  & 22 & 9\% & 22 & 9\% & 232 \\\hline
		Bonded IOBs  & 360 & 52\% & 360 & 52\% & 692 \\\hline
		CGLKs  & 8 & 50\% & 8 & 50\% & 16 \\\hline
		DCM  & 6 & 75\% & 6 & 75\% & 8 \\\hline
		\end{tabular}
	\label{tab:fsresources}
 }
\end{table}

Finally, to assess the complexity of our module in terms of resource usage, Table~\ref{tab:fsresources} highlights the increase in the resources between the reference switch design and the modified one that includes FairSwitch module.

\begin{figure}[t]
\captionsetup[subfigure]{justification=centering}
	\centering	 
		\begin{subfigure}[!ht]{0.52\columnwidth}
        \includegraphics[width=\textwidth]{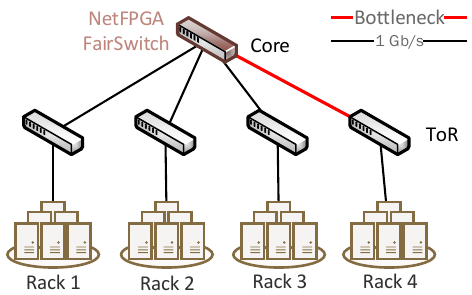}
	   \caption{The testbed topology}
	   \label{fig:topology}
    \end{subfigure}
	\begin{subfigure}[!ht]{0.4\columnwidth}
     \includegraphics[width=\textwidth]{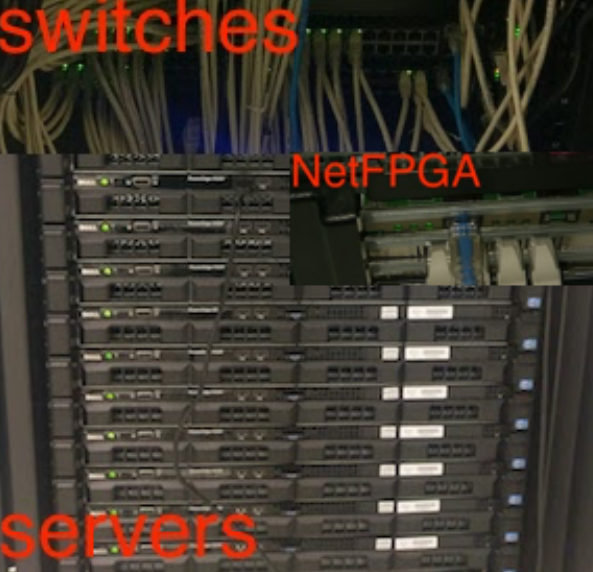}
	  \caption{The actual testbed}
	    \label{fig:actualtestbed}
       \end{subfigure}
	\caption{Testbed setup for FairSwitch evaluation}
	\label{fig:FStestbed}
\end{figure}

\textbf{Testbed Setup: } In the next set of experiments, we deploy the NetFPGA-based FairSwitch in a small-scale real-testbed. The testbed consists of 28 virtual servers, each server is associated with a physical dedicated 1 Gbps Network Interface Card (NIC). The serves are a set of high-performance Dell PowerEdge R320 machines. The machines are equipped with Intel Xenon E5-2430 6-cores CPU, 32 GBytes of RAM and Intel I350 server-grade 1 Gbps quad-port NIC. As shown in \figurename~\ref{fig:FStestbed}, the servers are organized into 4 racks (each rack is a subnet) and connected via 4 non-blocking Top-of-Rack (ToR) switches and the NetFPGA FairSwitch serves as the core switch of the network. The 4 racks are divided into (racks 1, 2 and 3) which are designated as senders and rack 4 which is designated as the receiver. Each 7 out of the 28 ports belonging to the same subnet is connected to one of the non-blocking ToR switches through 1 Gbps Ethernet links. The base RTT in the network is $\approx$200-300ms. The servers are installed with Ubuntu Server 14.04 LTS running kernel v3.18 which has by default the implementation of DCTCP~\cite{DCTCP}, Cubic and New-Reno (abbreviated to Reno) congestion control mechanisms. Finally, we also run on all end-hosts the FairSwitch end-host module which is implemented as a NetFilter-based loadable Linux kernel module~\cite{netfilter}. 

For experimentation purposes, the machines are installed with the iperf program \cite{iperf} for creating elephant flows and Apache benchmark for creating mice flows. We setup different scenarios to reproduce both ``incast" and ``incast and buffer-bloating" situations on the bottleneck link connected to the receiving rack 4. To emulate virtual guests and increase the number of senders dramatically, senders are attached to multiple virtual ports at the end-hosts on OvS. In this case, each iperf or Apache client/server process is directly associated with one of the virtual ports. This allows us to emulate traffic originating from any number of VMs and simplifies the creation of scenarios with a large number of flows in the network. The objectives of the experiments are: 
\begin{inparaenum}[\itshape i) \upshape]
\item to verify whether FairSwitch is able to support more TCP connections and maintains high link utilization; 
\item to verify the ability of FairSwitch to equally and fairly allocate bandwidth among conflicting mice and elephant TCP flows;
\item to verify how much FairSwitch is able to improve the FCT of the time-critical mice flows.
\end{inparaenum}

\begin{figure}[!ht]
\captionsetup[subfigure]{justification=centering}
\centering
	\begin{subfigure}[ht]{1\columnwidth}
           \includegraphics[width=\textwidth]{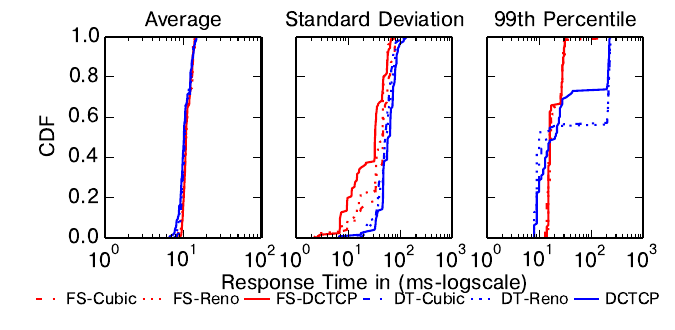}
						\caption{Medium incast scenario: 126 concurrent mice flows}
						\label{fig:126mice}
       \end{subfigure}
	\\
	\begin{subfigure}[ht]{1\columnwidth}
             \includegraphics[width=\textwidth]{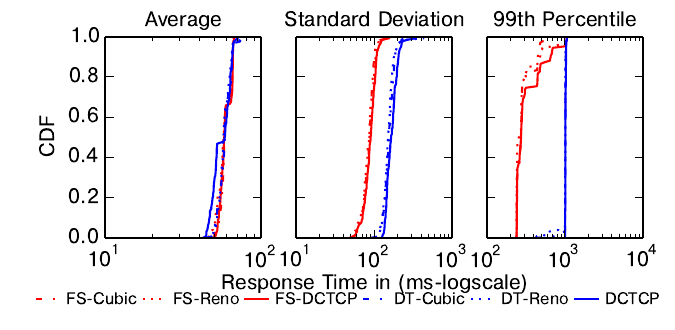}
						\caption{Heavy incast scenario: 630 concurrent mice flows}
              \label{fig:630mice}
       \end{subfigure}
				\caption{Incast Scenario: Average, Standard Deviation and Max FCT for TCP with FairSwitch vs DropTail vs DCTCP. Each flow requests 11.5MB file (divided into 1000 11.5KB blocks).}
				\label{fig:incast1}
\end{figure}

\subsection{NetFPGA Hardware-Switch Experimental Results}

\textbf{Incast Scenario without Background Traffic: } We run two mildly and heavily loaded incast scenarios where a large number of mice flows request a large content divided into 11.5KB chunks. 

\textbf{Experimental Setup: } In both scenarios, each of the 7 servers in rack 4, issues web requests via the Apache benchmark tool \cite{apacheb} requesting \textbf{"index.html"} page of size 11.5KB from each of the 21 servers in rack 1, 2 and 3. Hence a total of 126 ($21 \times 7 - 3 \times 7$) synchronized requests are issued. In the mildly loaded scenario, each request is repeated a thousand consecutive times by Apache benchmark which is equivalent to the transfer of an 11.5 MByte file for each requester ($11.5 \times 126 \approx$ 1.5 GBbytes total transfer through the bottleneck link). In the heavily loaded case, we repeat the same experiment with a thousand consecutive requests however, we use now five parallel TCP connections instead of just one. This is the same amount of transfer through the bottleneck link within the same period as in the mild load but with more concurrent connections. Note that, Apache benchmark, at the $0^{th}$sec, starts requesting the web page 1000 times then it reports different statistics over all the requests. 

\begin{figure}[!ht]
\captionsetup[subfigure]{justification=centering}
\centering
	\begin{subfigure}[ht]{0.48\columnwidth}
       \includegraphics[width=\textwidth]{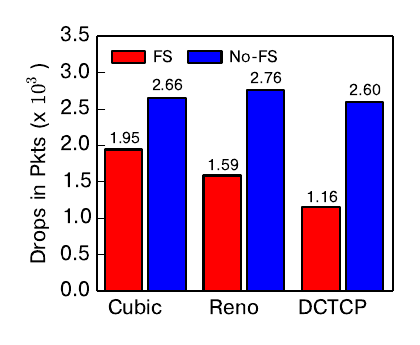}
                 \caption{The medium load case}
                \label{fig:mediumloss}
        \end{subfigure}				
		\hfill
		\begin{subfigure}[ht]{0.48\columnwidth}
       \includegraphics[width=\textwidth]{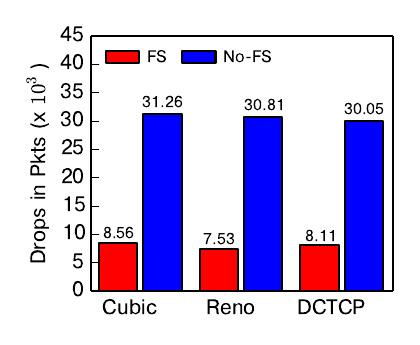}
                 \caption{The heavy load case}
                \label{fig:heavyloss}
        \end{subfigure}
				\caption{Incast Scenario: Total Packet Drops during the experiment for TCP with FairSwitch vs DropTail vs DCTCP.}
				\label{fig:incastdrop}
\end{figure}

\textbf{Experimental Results: }\figurename~\ref{fig:incast1} shows that, under heavy load, FairSwitch achieves a significantly improved performance for TCP flows in terms of the FCT variation from average and the FCT of the tail-end flows. Even though \scheme\! still improves both metrics in the mild case, the performance gains are not significant compared to the heavy scenario. The competing mice flows benefit under FairSwitch in the mild case by achieving almost the same FCT on average but with an order-of-magnitude smaller standard deviation compared to TCP (Cubic, Reno) with DropTail and DCTCP. In addition, it can improve the FCT of the tail-end (i.e., the maximum FCT) by two orders-of-magnitude suggesting that almost all flows (including tails) can meet their deadlines. Moreover, \figurename~\ref{fig:incast1} shows that more gains are obtained in the heavy load experiment thanks to the agility and fast convergence of \scheme\! scheme (or its FairSwitch deployment). Finally, the results suggest that FairSwitch can help incast traffic grab their fair-share quickly and thanks to the drop rate which is significantly reduced. \figurename~\ref{fig:incastdrop} shows that the number of drops during these incast experiments is reduced by $\approx55\%$~and~$\approx73\%$ in the mild and heavy load scenarios, respectively.

\textbf{Low frequency Incast with Background Traffic:} In the following experiments, we run a low frequency scenario where mice flows compete with elephants flows. Our goal is to see if FairSwitch can help mice flows grab some bandwidth from elephants and to see the effects on the throughput of elephants. 

\textbf{Experimental Setup: } We first generate 21 synchronized iperf \cite{iperf} elephant flows from each server in racks 1, 2 and 3 towards rack 4 through the bottleneck link continuously sending for 20 seconds. Then, we again invoke Apache-benchmark on the servers of rack 4 to request \textbf{"index.html"} from each of the web servers running on the servers in racks 1, 2 and 3. Hence, these web request must compete for the bottleneck bandwidth with each other and most importantly with iperf elephant traffic. After elephants have reached steady state (i.e., at the $10^{th}$ second), a single incast epoch consisting of 126 flows issue 100 consecutive web requests (i.e., each client requests a 1.15MB file partitioned into 100 11.5KB chunks) and then different statistics are reported. 

\begin{figure}[!ht]
\captionsetup[subfigure]{justification=centering}
\centering
	   \begin{subfigure}[ht]{1\columnwidth}
           \includegraphics[width=\textwidth]{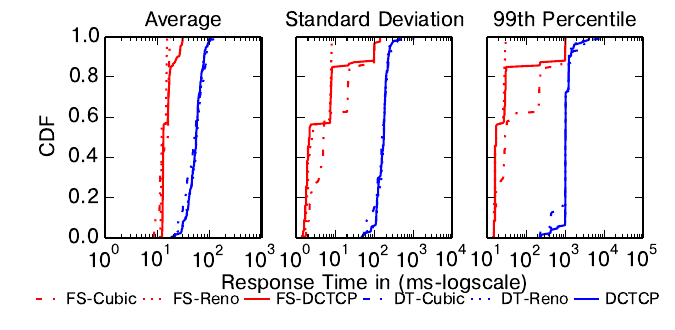}
						\caption{The various FCT metrics for one epoch of 126 concurrent mice flows}
						\label{fig:eleph126mice}
       \end{subfigure}
			\\
		\begin{subfigure}[ht]{0.48\columnwidth}
       \includegraphics[width=\textwidth]{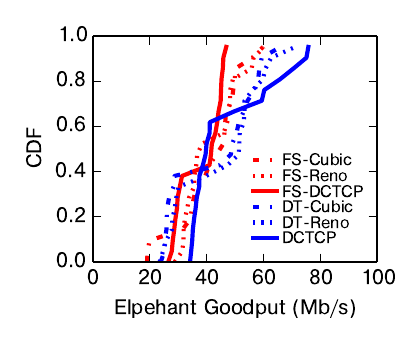}
                 \caption{Average elephant goodput}
                \label{fig:mediumgood}
        \end{subfigure}	
        \hfill			
	   \begin{subfigure}[ht]{0.48\columnwidth}
       \includegraphics[width=\textwidth]{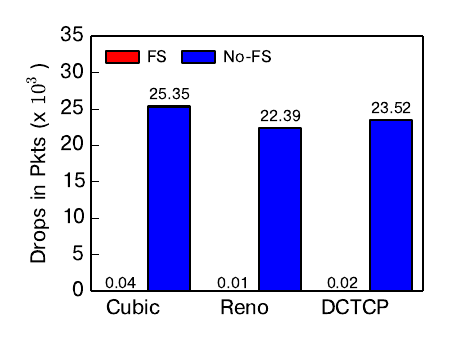}
                 \caption{Total packet drops}
                \label{fig:mediumlosswitheleph}
        \end{subfigure}
				\caption{Incast with background traffic: TCP with FairSwitch vs DropTail vs DCTCP. Each of the 126 mice flow requests once a 1.15 MB file (divided into 100 11.5KB blocks) while competing with 21 elephants.}
				\label{fig:incast-eleph1}
\end{figure}

\textbf{Experimental Results: }\figurename~\ref{fig:eleph126mice} shows that, in medium load, FairSwitch achieves FCT improvements for mice while nearly not affecting the performance of the elephants. Mice flows benefit with FairSwitch by improving the FCT on average and with one order-of-magnitude reduction in FCT variation compared to TCP (Cubic, Reno) with DropTail and DCTCP. Also, in terms of the tail-end (i.e., the last and similarly the $99\%$), FairSwitch reduces the tail FCT by two order-of-magnitude almost close to the average, and the FCT values are within 10's of ms. The improvement means mice flows finish quickly within their stipulated deadlines. Fig~\ref{fig:mediumgood} shows that the elephant flows are almost not affected by FairSwitch intervention and the throttling of their rates during the incast epochs. \figurename~\ref{fig:mediumlosswitheleph} shows that the drops under FairSwitch is reduced by up to $\approx99\%$ due to its efficient rate control during incast which explains why mice flows can avoid long waiting for timeouts.

\textbf{High frequency Incast with Background Traffic: } We repeat the above experiment, increasing the frequency of mice incast epochs to nine times within the 20 second period (i.e., at the $2^{nd}$, $4^{th}$, ..,  and $18^{th}$ sec). In each epoch, each server requests the web page 100 times (i.e., 1.15MB file partitioned into 100 11.5KB chunks). The total transfer is $\approx$ 145 MBytes per epoch and $\approx$ 1.3 GBytes for all 9 epochs. 

\begin{figure}[!ht]
\captionsetup[subfigure]{justification=centering}
\centering
			\begin{subfigure}[ht]{1\columnwidth}
           \includegraphics[width=\textwidth]{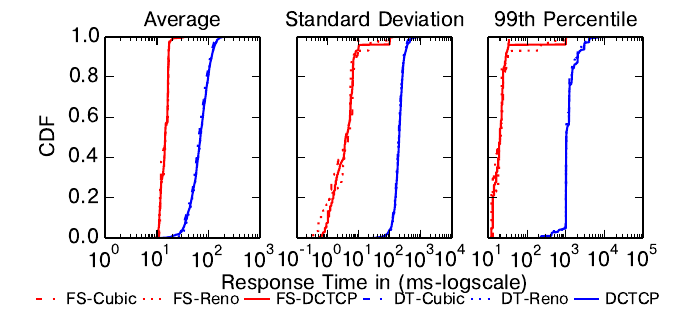}
						\caption{The various FCT metrics for 9 epochs of 126 concurrent mice flows}
						\label{fig:high126mice}
       \end{subfigure}
			\\
			\begin{subfigure}[ht]{0.48\columnwidth}
       \includegraphics[width=\textwidth]{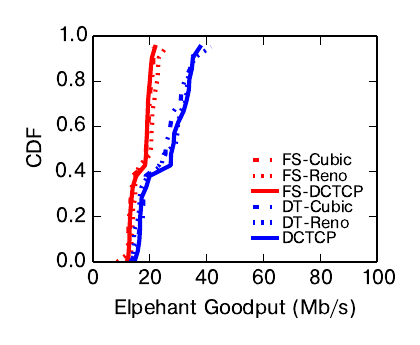}
          \caption{Average elephant goodput}
           \label{fig:highgood}
        \end{subfigure}		
        \hfill		
		\begin{subfigure}[ht]{0.48\columnwidth}
       \includegraphics[width=\textwidth]{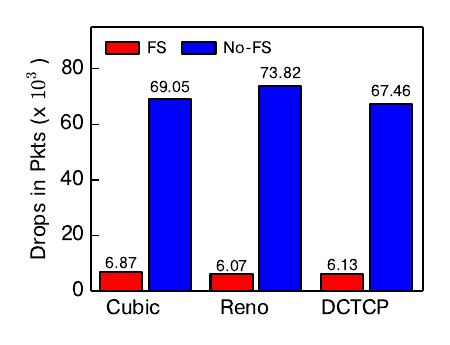}
                 \caption{Total packet drops}
                \label{fig:highloss}
        \end{subfigure}
				\caption{Incast with background traffic: TCP with FairSwitch vs DropTail vs DCTCP. Each of the 126 mice flow requests 9 times a 1.15 MB file (divided into 100 11.5KB blocks) while competing with 21 elephants}
				\label{fig:incast-eleph2}
\end{figure}

\textbf{Experimental Results:} \figurename~\ref{fig:high126mice}, even with the higher incast frequency, FairSwitch is still able to keep up with the higher incast frequency even when mice flows are fighting their way against fat elephant flows. The average and variation of FCT for mice flow see similar improvement as the previous experiment. This can be attributed to the lower packet drop rate of up to $\approx92\%$ with the help of FairSwitch and hence lower chances of experiencing timeouts as shown in \figurename~\ref{fig:highloss}. Compared to the previous experiment, \figurename~\ref{fig:highgood} shows that elephants' throughput is slightly reduced because of the equal rate allocation for a large number of incast flows and the few  elephants during the incast epochs. However, we believe that the fair utilization of the bandwidth by mice and elephants during incasts is necessary for mice flows to finish in a timely manner which explains the lower achieved goodput by elephants.

In summary, all the experimental results of the kernel module, Open vSwitch and hardware prototypes support and highlight the performance gains, especially for OLDI applications obtained by adopting \scheme\! system. In particular, they show that:
\begin{itemize}
\item It reduces the FCT variance and tail-end FCT for mice flows by up to two orders-of-magnitude. 
\item It can maintain the same improvement even if the bandwidth-hungry elephants are hogging the network.
\item It efficiently handles incast, in low and high frequency, by allowing mice to compete in fairly with elephants.
\item It fulfils its requirements with no more than default assumptions about the network stack and without imposing any modifications to tenants' VMs.
\end{itemize}

\section{Related Work}
\label{sec:related}

Due to the impact and severity of the aforementioned congestion symptoms, much recent work has been devoted to addressing such shortcomings of TCP in DCNs. Some approaches explored flow path scheduling schemes to isolate mice and elephants~\cite{Wenxue2016,Wei2016}. For instance, Freeway~\cite{Wei2016} leverages path diversity in the DCN topology to improve mice flow completion time. However, these schemes require estimation of path delays and solving optimization problems that are not suitable for time-scales of latency in data centers.  Instead, most of the proposed solutions fall into two categories: window-based schemes (e.g., \cite{Ahmed-LCN-2016,Ahmed-LCN-2017,Alizadeh2010,Wu2013, Ahmed-GLOBECOM-2018,Ahmed-GLOBECOM-2015,Ahmed-INFOCOM-2019,Ahmed_ToN_2023}) or fast loss recovery schemes (e.g., \cite{Vasudevan2009, Cheng2014, Ahmed-INFOCOM-2018, Ahmed-ICDCS-2019-1, Ahmed-ICPP-2020}). 

In the window-based category, DCTCP \cite{Alizadeh2010} proposes a modification to TCP and RED active queue management that adjusts TCP's congestion window to stabilize the queue length in the switch at a predefined small threshold, guaranteeing thus short delays for incast traffic, without degrading the link utilization. ICTCP \cite{Wu2013} also was proposed as a modification to TCP receiver to handle incast traffic. ICTCP adjusts the TCP receiver window proactively, to avoid congestion at the receiver. The experiments with ICTCP in a real testbed show that ICTCP can almost curb timeout-detected losses and achieves a high throughput for TCP incast traffic, however, since it is focused on incast it only handles congestion at the receiver and does not address buffer buildup in the switches. 

Fast loss recovery schemes try to improve the agility of TCP in recovering from congestion events by shortening the reaction time. For instance, \cite{Vasudevan2009,Ahmed-INFOCOM-2018} reduces TCP's minimum retransmission timeout $RTO_{min}$ to reduce the unnecessarily long waiting times after packet losses to enable a fast reaction to congestion losses in the presence of shallow buffers (where losses are mostly detected by timeout). In contrast \cite{Cheng2014} cleverly tries to deploy a fast congestion-detection mechanism by truncating the packet payload of congestion-causing packets, only conveying the header to the receiver. This enables a receiver-driven explicit congestion-notification upon reception of truncated packets. Fast loss recovery schemes potentially solve the problems of congestion in data centers, however, they require not only switch modification, but also end-system modifications. For example, in Linux $RTO_{min}$ is equal to 200ms and is hard-coded in the TCP source code. Other approaches consider the co-flow abstraction to collectively optimize the performance of flows who share the same goal or task~\cite{Ahmed-ICDCS-2019-2,Ahmed-SRDS-2017}.

Alternatively, we have explored an end-to-end flow-aware approach~\cite{Ahmed-CLOUDNET-2015,Ahmed-IPCCC-2015} leveraging explicit feedback from in-network devices similar to traditional flow-based systems like ATM-ABR~\cite{ATM-ABR} or XCP~\cite{Katabi2002}. We have addressed a practically difficult challenge: how to deploy such flow-awareness in the flow-averse IP environment without modifying the TCP sender and receiver. This disqualifies XCP, as it is a clean-slate redesign that requires not only changes to the routers but also to the sender and receiver. To achieve our goal, the switch/router is set to track flows, calculate a fair share for each flow that traverses it, and convey back this fair share to the source. In our approach, we enable low-profile flow awareness and modify the switch software to rewrite TCP receiver window to communicate with the sender. Hence, our proposed mechanism fits in well for data centers without any change to TCP at the end-hosts. These works were followed up with SDN-based designs which leverage the same idea towards solving the problem~\cite{Ahmed-ICC-2016-1,Ahmed-ICC-2016-2,Ahmed-LCN-2017}.

TCP optimization plays a pivotal role in the efficiency and reliability of federated learning (FL) protocols by influencing communication between devices~\cite{Ahmed-IoTJ-2023,REFL_2023}. FL, often executed over wireless networks, relies on efficient data transmission among devices. TCP optimization strategies, such as adjusting congestion control algorithms or packet loss recovery mechanisms, become crucial~\cite{Ahmed-AQFL-21,Amna-FedEdge-2022}. Optimizing TCP for FL ensures faster convergence of models, minimizes communication delays, and enhances the robustness of data transmission across distributed devices. This optimization directly impacts the overall performance and success of FL by mitigating communication bottlenecks, ensuring consistent and reliable data transfer, and ultimately facilitating more effective collaborative learning among decentralized devices.

\section{Conclusion}
\label{sec:conclude}
In this work, we explore a non-intrusive way of reconciling between the long-lived (elephants) and short-lived (mices) flows. To achieve this, the persistent switch queue sizes should operate at low levels to make room for the bursts of incast traffic which helps avoid packet-losses. We proposed, \scheme\!, a switch-assisted flow-aware rate matching algorithm that only relies on the existing flow-control mechanism of TCP to feedback queue occupancy levels to TCP senders. To show the practicality of \scheme\!, we provide prototypes implemented as Linux-Kernel module for bare-metal switches, OpenvSwitch patch for virtual DC networks and NetFPGA module for programmable hardware switches in data centers. A number of detailed simulations and real test-bed experiments showed that \scheme\! can achieve its goals efficiently while outperforming the most prominent alternative approaches. Last but not least, knowing that in most public data centers the TCP sender and/or receiver are outside the control of the DCN operator, \scheme\! makes a point of principle to not modify the source TCP congestion control algorithms to enable true deployment potential in real public DC networks.
\bibliographystyle{ieeetr}
\bibliography{refs,self,mypapers,online}%

\end{document}